# Equivalent circuit model for electrosorption with redox active materials

Fan He[1], Martin Z. Bazant[1,2], T. Alan Hatton[1]*

[1]Department of Chemical Engineering, Massachusetts Institute of Technology, Cambridge, Massachusetts 02139, USA.

[2]Department of Mathematics, Massachusetts Institute of Technology, Cambridge, Massachusetts 02139, USA

* To whom correspondence should be addressed. E-mail: tahatton@mit.edu

## Abstract

Electrosorption is a promising technique for brackish water deionization and waste water remediation. Faradaic materials with redox activity have recently been shown to enhance both the adsorption capacity and the selectivity of electrosorption processes. Development of the theory of electrosorption with redox active materials can provide a fundamental understanding of the electrosorption mechanism and a means to extract material properties from small-scale experiments for process optimization and scale-up. Here, we present an intuitive, physics-based equivalent circuit model to describe the electrosorption performance of redox active materials, which is able to accurately fit experimental cyclic voltammetry measurements. The model can serve as an efficient and easy-to-implement tool to evaluate properties of redox active materials and help to distinguish between the transport-limited and reaction-limited regimes in electrosorption processes. And the extracted intrinsic material properties can be further incorporated into process models under lower supporting electrolyte concentrations for realistic electrosorption applications.

**Keywords**: Theory of electrosorption, Redox active electrodes, Electrical double layer, Electrochemical adsorption, Environmental remediation

## Introduction

Electrosorption with redox active (or Faradaic) electrode materials has been a rapidly growing area of research and development over the past few years, and has shown promise for applications in low concentration brackish water treatment and waste water remediation [1,2]. Unlike traditional capacitive deionization (CDI), which works by electro-sorption of ionic species within ideally polarizable (blocking, non-reactive) porous electrodes through the formation of electric double layers [3–5], electrosorption processes with Faradaic materials involve electrochemical half-cell reactions at redox active electrodes, such as specific intermolecular interactions with polymeric redox moieties (electrochemically mediated adsorption, EMA) [6–10] or solid-state ion intercalation (desalination batteries) [11–15]. Since both the EMA and intercalation electrodes exhibit pseudo-capacitance controlled by Faradaic reactions, such redox active materials used for electrosorption will be referred to as "pseudo-capacitive" or "Faradaic" materials in this paper.

Theory development has progressed along with innovations in electrosorption materials and device configurations for water treatment applications. Over the past decade, in particular, models with hierarchical complexities have been developed to describe the electrosorption behavior ranging from the double-layer (microscopic) to the device (macroscopic) level. A relevant question is what the proper model for the problem of interest might be. Existing models based on one-dimensional and two-dimensional numerical frameworks can be complicated to implement and require expertise in numerical computing; furthermore, depending on the nature of the problem, spatial dependence of the variables may not be significant. For these reasons, zero-dimensional models are of value for direct comparison with experimental measurements. In this paper, following a brief review of the published electrosorption models, we propose a simple equivalent-circuit model that incorporates both thermodynamic and kinetic effects in Faradaic electrosorption processes and extract appropriate physical parameter values from fits of the model to experimentally measured cyclic voltammograms for porous electrodes with redox-active materials.

## History of Electrosorption Models

Traditional CDI is based on the electrostatically driven adsorption of ions in the electric double layer (EDL) of a blocking (non-reactive) metal electrode. EDL theory was introduced by Helmholtz in 1853 [16] to describe a monolayer of counter-ions in solution screening surface charge like a capacitor. In the early 20$^{th}$ century, Gouy recognized that the screening layer in the electrolyte is often diffuse, containing both counter-ions and co-ions spread over larger distances, and developed the original mean-field theory of EDL structure based on the Poisson-Boltzmann equation [17]. Chapman then derived the EDL differential capacitance for Gouy's model for the case of a symmetric binary electrolyte [18], and Stern introduced a series capacitance representing a surface solvation layer to correct for the unrealistically diverging capacitance predicted at high voltages [19]. In the ensuing century, many more sophisticated double-layer models have been developed for high voltages or concentrated solutions [20], but the Gouy-Chapman-Stern (GCS) model remains widely used to this day for its simplicity and its validity in the dilute regime, relevant for desalination processes.

The first engineering model of CDI was developed by Johnson and Newman in 1971 [21], assuming only counter-ion adsorption in an equivalent circuit model for EDL charging. A linear diffusion equation, based on de Levie's RC transmission line model for porous electrodes with constant EDL capacitance and constant electrolyte conductivity per cross sectional area [22], is solved for the volume-averaged EDL charge profile, which acts as a sink for counter-ions in the ambipolar diffusion equation for a neutral binary electrolyte. This seemingly reasonable approach, however, is inconsistent with the GCS model since it neglects co-ion desorption and fails to capture the nonlinear coupling of salt diffusion, ion adsorption and charge relaxation.

The first self-consistent, microscopic theory of salt depletion by EDL charging was developed by Bazant, Thornton and Adjari in 2004 [23], albeit motivated by induced-charge electrokinetic phenomena [20,24] rather than by capacitive deionization (CDI) or water treatment. They analyzed the Poisson-Nernst-Planck (PNP) transport equations for a dilute, binary electrolyte between parallel blocking electrodes and identified three dynamical regimes for the response to an applied voltage: (i) the classical RC circuit model corresponds to linear response to small applied voltages with zero net salt adsorption by the EDLs, as counter-ion attraction is balanced by co-ion repulsion from the electrodes; (ii) for thin EDLs (small Debye screening length) compared to the electrode spacing, the "weakly nonlinear" dynamics corresponds to quasi-equilibrium EDLs having the differential capacitance of the GCS model in series with the bulk electrolyte resistance, still without concentration gradients; (iii) salt adsorption from the bulk solution by the EDLs only occurs in the "strongly nonlinear" regime, where net counter-ion adsorption exceeds co-ion desorption from the two electrodes [23]. The same three dynamical regimes arise for arbitrary electrical forcing, such as large alternating current [25], and in arbitrary geometries, including ideally polarizable particles [26] and porous electrodes [27] as we now explain.

The self-consistent, macroscopic theory of electrosorption processes began a decade ago, as surveyed in the timeline summarized in
Figure 1 by equivalent circuit elements representing the interfacial processes considered. Biesheuvel and Bazant [27] extended the thin-double layer approximations of Bazant, Thornton and Ajdari to blocking porous electrodes by formal volume averaging and thus derived the first mathematical model of CDI, including both counter-ion adsorption and co-ion repulsion. This analysis established a bridge between EDL theory and realistic applications in water treatment and desalination by capacitive deionization using porous carbon electrodes. The linear and weakly nonlinear regimes were shown to correspond to an RC transmission line of solution resistance and EDL capacitance, while salt adsorption/desorption cycles correspond to the strongly nonlinear regime, which cannot be described by an equivalent circuit model. Mirzadeh, Gibou and Squires later validated the thin-EDL model against direct numerical simulation of the PNP equations in porous electrodes and proposed corrections for surface conduction [28].

A number of extensions of the model soon followed. Biesheuvel, Fu and Bazant [29,30] added the generalized Frumkin-Butler-Volmer model [31,32] for parasitic Faradaic reactions, such as metal electrodeposition/dissolution or water splitting, consistent with the GCS model of the EDL, but they did not account for any pseudo-capacitance resulting from reversible charge storage via such reactions. In the linear and weakly nonlinear regimes, the model was shown to correspond to an RC transmission line with shunt resistors around the double layer capacitors. The same work also introduced the "modified Donnan model" for thick double layers in a hierarchical porous electrode, where charge is stored throughout a "micropore" volume, coupled with ion transport through connected "macropores" [29]. Biesheuvel and collaborators later extended this model for attractive image forces [33] as well as fixed chemical charge on the electrode surface within the EDL, which could either enhance or retard desalination, depending on the point of zero charge [34,35].

Recently, redox active materials have been applied for electrosorptive water remediation, where the EDL capacitance is considered negligible compared to the pseudo-capacitance resulting from the reversible Faradaic reactions. The categories of redox active materials that have been studied so far include but not limited to intercalation materials as well as redox active polymers. And there has been parallel theoretical development for both categories. In 2017, Smith started the theoretical evaluation of using ion intercalation materials for desalination application, where he has used a lattice-based ideal solid solution model to describe the pseudo-capacitance of galvanostatic charge/discharge of nickel hexacynoferrate (NiHCF) at low C-rate in 1 M $Na_2SO_4$, assuming fast ion intercalation reactions [36]. Soon after, Porada et al. [37] analyzed the open circuit voltage (OCV) profile of NiHCF determined by galvanostatic intermittent titration and found that the data were well described by the Frumkin isotherm [38,39], which is equivalent to the regular solution model used in phase-field models of Li-ion battery materials [40]. Based on this approach, Singh et al [41] developed a general theory of water desalination by electrosorption with intercalation materials, similar to Li-ion battery models [42] except that they assumed fast reactions and neglected Faradaic charge-transfer resistance. On the other hand, in 2018, He et al [43] published the first model to take into account the thermodynamics of electrosorption processes enhanced by potential dependent surface chemical charges, which is described by a variable Faradaic capacitance coupled with the GCS double layer capacitance. A Faradaic charge transfer resistance is introduced in the same work, to account for more realistic charge transfer kinetics from the electrode surface to the redox active species immobilized within the pores of the sorbent. Till now, there has been no published effort in unifying the theoretical descriptions of redox active materials. This work is also our attempt to develop a generic and simple theoretical frameworks for analyzing and extracting material properties of redox active electrodes from electroanalytical experimental data.

## Electrosorption Circuit Model for Redox Active Materials

In general electrosorption processes, both the thermodynamics and kinetics of Faradaic reactions must be considered. Here, we have adopted the same approach from the earlier publication [43] to give a physics-based equivalent circuit model that can represent the electroanalytical performance of Faradaic materials. In this context, we are particularly interested in the studying the supporting electrolyte under high concentrations, as is typical in cyclic voltammetry experiments, where there is no significant depletion of the bulk electrolyte, mass transfer limitations on the migration of the mobile ionic species can be minimized and focus can be placed on the charge transfer behavior of the redox active material itself. The intrinsic material properties extracted by fitting the model to experimental data can be further incorporated into more detailed transport model or lumped parameter process-level models [43] under lower supporting electrolyte concentrations for realistic water treatment applications, where the ion concentrations can vary in both space and time. The EDL capacitance is included only as a circuit element for linear response, while the strongly nonlinear dynamics of EDL salt removal are neglected compared to effects of pseudocapacitance from the Faradaic reactions. The same approximations are also made in porous electrode theories of intercalation batteries [38,42,44,45], so we will not dwell further on their justification here.

A great benefit of equivalent circuit models is that they are simple and intuitive enough to interpret, because each circuit element has a clear physical meaning and can be well characterized and verified by experimental measurement, such as impedance spectroscopy, chronopotentiometry, etc. The equivalent circuit model we developed here to describe the electrochemical charging and discharging process in redox-active materials is shown in Figure 2, where $C_F$ denotes the Faradaic (pseudo-)capacitance, $R_F$ denotes the Faradaic charge transfer resistance, $R_s$ is the solution resistance, $C_c$ denotes the double layer capacitance, and $R_l$ is the leakage resistance from parasitic reactions. In contrast to typical equivalent circuit models where the value of each element is assumed to be constant (invariant of local electric potential), however, each of the dominant circuit elements in our model is described by the underlying physics, and therefore the model is expected to generalize better to a wide range of experimental conditions than are traditional simple equivalent circuit models.

In the system in which we are interested, the redox-active moieties are mostly uncharged initially, but undergo a fast, reversible, one-electron transfer reaction [46] upon application of a positive potential such that their reduced state is oxidized to the positive charged state, such as the conversion of ferrocene to ferrocenium on a redox polymer polyvinylferrocene; an anion is then adsorbed near the redox moiety through electrostatic, intermolecular hydrogen bonding and other interactions to balance the charge [7]. The electrochemical reaction for the electrosorption process is written as,

$$R - e^- + A^- \rightleftharpoons O^+ A^- \tag{1}$$

where $R$ and $O^+$ refer to the reduced and oxidized forms of the redox moieties, respectively, and $A^-$ is the targeted anion species in solution.

Each of the circuit elements can be described in terms of physically identified variables, as shown below.

The equilibrium voltage (i.e. OCV) is governed by the Frumkin isotherm (regular solution model) [37,39,41].

$$\Delta\phi_{eq} = \Delta\phi_{eq}^\Theta + \frac{RT}{F}\ln\left(\frac{\theta_O}{1-\theta_O}\right) + \frac{RT}{F}\ln\left(\frac{c_{A^-}}{c^\Theta}\right) + \Omega(1 - 2\theta_O) \tag{2}$$

where $\theta_O$ denotes the surface coverage of the oxidized state or, equivalently, the state of charge (SOC) of the redox-active electrode (the oxidized ferrocenium possesses a net positive charge of one); $c_{A^-}$ is the concentration of anions in the solution, and $c^\Theta = 1$ M as the standard reference concentration. The last term takes into account the enthalpic interactions between the charged and uncharged states, where $\Omega > 0$ indicates a repulsive force within the charged states and an attractive force between the charged and uncharged states, which inhibit phase separation [40]. Note that $\Omega$ is equivalent to the parameter $g$ in reference [37,41] with the following relation $\widetilde{\Omega} = \frac{\Omega}{RT} = \frac{1}{2}\tilde{g} = \frac{eg}{2k_BT}$.

When the enthalpic interaction is not important, equation (2) results in the familiar expression for the Nernst equation with a Langmuir isotherm (ideal solution model) [36].

When the bulk concentration of the supporting electrolyte is identical to the reference concentration and when the enthalpic interaction is negligible, the last two terms of equation (2) disappear and the pseudo-capacitance $C_F$ can be expressed as a function of the local potential difference under equilibrium condition [47], as follows,

$$C_F = \frac{dQ}{dV} = \frac{F\Gamma_s d\theta_O}{d\Delta\phi_{eq}} = \frac{F^2\Gamma_s}{4RT}\text{sech}^2\left(\frac{F(\Delta\phi_{eq} - \Delta\phi_{eq}^\Theta)}{2RT}\right) \tag{3}$$

where $\Gamma_s$ denotes the total surface site density of the redox active species, in unit of mol m$^{-2}$.

The kinetics of the electron transfer reaction are described by the Butler-Volmer equation,

$$i_F = i_0 \left[\exp\left(\frac{F(1-\alpha)\eta}{RT}\right) - \exp\left(-\frac{F\alpha\eta}{RT}\right)\right] \quad (4)$$

where we adopt the exchange current density derived from non-equilibrium thermodynamics by Bazant [40] for a uniform surface concentration, assuming Frumkin/regular-solution thermodynamics with one excluded surface site in the transition state of the reaction:

$$i_0 = k_0 F \Gamma_s \theta_R^\alpha \theta_O^{1-\alpha} a_A^\alpha e^{(1-\alpha)\widetilde{\Omega}(1-2\theta_O)} \quad (5)$$

where $\alpha$ is the charge transfer coefficient for the single electron transfer reaction, and $\eta = \Delta\phi_{eq} - \Delta\phi_{eq}^\Theta$ is the overpotential that drives the Faradaic reaction. The activity of the anions is defined as $a_{A^-} = c_{A^-}/c^\Theta$.

The Faradaic charge transfer resistance is defined as

$$R_F = \eta/i_F \quad (6)$$

Under small overpotentials, equation (4) can be linearized to give $i_F = i_0 F\eta/RT$, and under these conditions, equation (6) indicates that the Faradaic resistance is inversely proportional to the exchange current density, *i.e.*, $R_F \propto 1/i_0$.

The rate of change of the state of charge, or equivalently, the surface concentration of the oxidized state is linked to the Faradaic current corresponding to the charge transfer reaction,

$$\frac{d\theta_O}{dt} = \frac{i_F}{\Gamma_s F} \quad (7)$$

Kirchhoff's current law can be used to relate the measured total current to the currents within the individual pathways shown in Figure 2, *i.e.*,

$$I = I_F + I_c + I_l \quad (8)$$

Where $I_F = Ai_F$, $I_l = (\Delta\phi_{eq} + \eta)/R_l$, $I_c = \nu C_c$, $A$ is the nominal surface area of the electrode and $\nu$ is the voltage sweep rate, in unit of V/s.

The applied voltage is simply the combined voltage drop across the solution resistance, the pseudo-capacitor and the charge transfer resistor for the Faradaic reaction,

$$V = IR_s + \Delta\phi_{eq} + \eta \quad (9)$$

The applied voltage in any half-cycle in a cyclic voltammetry experiment can be determined from,

$$\frac{dV}{dt} = \pm \nu \quad (10)$$

Therefore, $V = V_0 \pm v(t - t_0)$, where the positive and negative signs are for the oxidation and reduction steps, respectively, and $V_0$ denotes the potential at the beginning $t_0$ of a given cycle.

## Experiments

There is a wide variety of redox active materials that can be used for electrosorption applications, including conducting polymers, redox polymers, and intercalation compounds [2,48]; in this study we used polyvinylferrocene (PVFc), a redox polymer with fast, reversible, one-electron transfer kinetics [46]. Redox activity within this redox polymer is modulated by an electron-hopping mechanism between ferrocene moieties since the polymer backbone does not exhibit high electronic conductivity, but it has been shown by Mao et al [49] that the electron transfer kinetics can be greatly enhanced by attachment through π-π interactions of the PVFc to multi-walled carbon nanotubes that act as molecular wires to facilitate electron transfer within the polymer layer.

**Materials**

Polyvinylferrocene (PVFc) was purchased from Polysciences. Multi-wall carbon nanotubes (CNT) and sodium perchlorate were obtained from Sigma Aldrich. All chemicals were used as received without further purification. Toray carbon paper with no PTFE treatment (TGP-H-60) was purchased from VWR Scientific. Copper wire and copper tapes were supplied by McMaster.

**Electrode preparation**

Electrode substrates were prepared by wrapping 1 cm by 3 cm swatches of carbon paper with copper tape and wire as the current collector. Epoxy was used to mask the exposed surface area of the electrode to the electrolyte to be 1 cm by 1 cm.

To prepare the PVFc/CNT hybrids, we followed the preparation procedure outlined by Su et al [7], which has been demonstrated to produce electrodes with good redox activities for electrosorption applications. Briefly, the PVFc/CNT electrodes were prepared by the drop casting of a mixture solution onto the carbon paper substrate. A stock solution A of 80 mg poly(vinyl)ferrocene (PVFc) and 40 mg multiwalled-CNT was dissolved in 10 mL anhydrous chloroform. A stock solution B of 40 mg of CNT in 10 mL chloroform was also prepared. The two stock solutions were sonicated for 1.5 hours in icy water to dissolve the polymer and disperse the CNTs, respectively. The PVFc/CNT ink was prepared by mixing stocks A and B in 1:1 ratio to obtain a 50%/50% by mass suspension of the PVFc/CNT complex. The mixture ink was sonicated for another 1.5 hour in an ice-bath to completely disperse the redox active polymer with CNTs. The ink was then drop-cast onto the carbon paper substrate by 50 μL (equivalent to 1 μmole of ferrocene units) and left to dry at room temperature.

**Cyclic voltammetry**

All cyclic voltammetric electrochemical studies were performed at room temperature on a Gamry Interface 1000E potentiostat in a 10 mL BASi MCA cell in the three-electrode configuration. A platinum wire was used as the auxiliary electrode and Ag/AgCl (3 M NaCl) as the reference electrode (BASi), with a 1 M NaClO4 aqueous solution as the electrolyte. Automatic *IR* compensation was used to correct for the solution resistance during the measurement, for which no prior knowledge was required; this protocol produced a consistent voltage window on the electrode-electrolyte interface across all scan rates throughout the measurement.

## Results and discussion

The physics based equivalent circuit model was solved numerically by the DAE solver ode15s in MATLAB 2018a. A triangular voltage profile was used as input to simulate the response of the redox-active materials under cyclic voltammetry. Figure 3(a) shows the calculated current response and Figure 3(b) the dynamics of the oxidation of the redox-active moieties (or the time-dependent state of charge) on the electrode surface under a certain scan rate. The current response matches well qualitatively with the typical features of redox active electrodes [1,9,49], where there is a pair of distinct peaks near the equilibrium voltage of the redox active material. The anodic and cathodic peak current intensities are identical but with opposite sign, indicating the reversibility of the Faradaic reaction. Further away from the peaks, the current decays to a plateau of near zero. These peaks and the diminishing current away from the equilibrium voltage distinguish redox active (pseudo-capacitive) electrodes from purely capacitive materials, where cyclic voltammetry usually shows a quasi-rectangular shape [50].

The fraction of redox moieties in the oxidized state is similar to the state of charge (SOC) in energy storage devices such as batteries. The different pathways for SOC development with applied voltage show hysteresis during the cyclic charging and discharging process, and are primarily due to the irreversible Faradaic charge-transfer resistance, without which the cyclic voltammetry would be perfectly symmetrical for the forward and backward sweeps.

As shown in Figure 4, the simulation results of the charging/discharging curves for the redox active materials from the equivalent circuit model at 100 mV/s (from Figure 3) are further compared with the previous studies from Porada et al [37], where the intercalation/deintercalation characteristics of an ion-intercalation material, nickel hexacynoferrate (NiHCF), is demonstrated. This is another important category of electrode materials with Faradaic reactions for deionization applications. The material was tested by the Galvanostatic intermittent titration technique (GITT) with a period of constant current charging/discharging followed by an open circuit step to allow relaxation of the Faradaic charge transfer resistance to the equilibrium voltage at any given state of charge between the charging steps. And the voltage at the end of each "rest" period was used to construct the equilibrium voltage versus electrode charge curves. The rest period between each charging/discharging steps avoided the hysteresis phenomena associated with the irreversible charge transfer resistance in their system as opposed to the prediction from the equivalent circuit model with $R_F$. The elimination of the hysteresis effect was confirmed by the excellent reversibility of the intercalation and deintercalation of sodium ions. And both of the equilibrium voltage curves were described perfectly by the Temkin (Frumkin) isotherm, equivalent to equation (2), which is a widely used correction for porous electrode modeling of sodium ion batteries. Although the effect of Faradaic charging resistance is important in dynamic processes and is evident from the previous studies with intercalation materials [36,37,41], to date it has not been studied quantitively in the literature in electrosorption theories.

Cyclic voltammetry responses can be further simulated under different scan rates to compare with the experimental data. The solution resistance $R_s$ is close to zero due to automatic *IR* compensation from the control loop in the hardware, as noted in the experimental section. Due to parasitic reactions are insignificant in this study, $R_l$ is fixed at a large constant [51]. And the model fitting parameters are $k_0, C_c, \Omega$, and $\Gamma_s$, which denote the charge transfer rate (in exchange current density) of the Faradaic reaction, double layer capacitance, the pair-wise interaction energy between the charged and uncharged states and the accessible total surface site density.

The simulation results shown in Figure 5(b) match well qualitatively with the experimental data. Both the simulation results and the experimental data show an increase in current intensity with scan rate in both the anodic and cathodic directions. Also, with higher scan rates the voltage at the peak current moves outwards relative to the equilibrium voltage. In the fitting of the physics-based equivalent circuit model to experimental data under various scan rates, the three parameters $k_0, C_c, \Omega$ are assumed to be unaffected by the scan rate, but $\Gamma_s$ is fitted to each individual data set under various scan rates. The best fit results (in a least-squares sense) are: $k_0 = 0.4 \text{ s}^{-1}$, $C_c = 50 \text{ F/m}^2$, and $\Omega = -620$ J/mol (i.e., $-0.0064$ eV). The simulation results for the redox active electrodes using the fitted parameters are overlaid with the experimental data at various scan rates in Figure 6. It can be seen that the model matches the experimental measurements well for medium to high scan rate ($v \geq 50$mV/s), where the electrodes exhibited more symmetric redox activities during the anodic and cathodic polarizations. Under slow scan rates ($v \leq 20$mV/s), the anodic peak was sharper than the cathodic peak, possibly due to polymer reconfiguration when solvent molecules and ions are doped into or undoped from the polymer matrix to balance the excess charge on the electrode [46].

It may be necessary to consider more sophisticated models of Faradaic reaction kinetics in future work. In all cases studied here, the simulated current after the redox peaks decayed more rapidly than the experimental data, which may be attributed to the unphysical unbounded exponential increase of current (and decrease of activation overpotential) in the phenomenological Butler-Volmer equation used here. In contrast, microscopic theories of electron transfer, pioneered by Marcus [52], such as the Marcus-Hush-Chidsey [53,54] or asymmetric Marcus-Hush [55] models of Faradaic reactions at metal electrodes, predict saturation to a reaction-limited current at large overpotentials, which may help to fit the data. Indeed, there is growing evidence that electron-transfer theory applies to the common Faradaic materials for electrosorption processes, including ferrocene-terminated functional groups on metal electrodes [53] as well as ion insertion compounds [42,56,57].

The extracted $\Gamma_s$ values are shown as a function of scan rate in Figure 7. The total accessible surface site density decreased with increasing scan rate, indicating an electron transfer resistance between the current collector and the electrode surface. The effect of that can be captured by a transmission line model with spatially distributed imperfect electronic conduction resistance [51] and is ignored to keep the simplicity of the model here.

The experimental and simulated anodic and cathodic peak currents are shown to be linear functions of scan rate ($R^2$>0.99 for experimental data) in Figure 8. This linearity is quite interesting, since in past literature peak currents of redox active electrodes were reported to vary with scan rate to the 1/2 power [58,59], attributed to the diffusional transport of the electroactive species to the electrode surface in accord with the well-known Randles-Sevcik equation for cyclic voltammetry [60]. However, as discussed earlier, the current response of the redox-active materials is mainly due to the Faradaic charge transfer resistance and the voltage dependent density of oxidation states (or SOC), the current response is not likely to be limited by the diffusion of mobile species in the solution, which begins and stays at a high concentration (i.e. 1 M) during the cyclic voltammetry measurements. To be more precise, the thickness of the redox active polymer was estimated to be $L = 20$ μm from cross sectional SEM measurements of the electrode under a loading similar to that used in this study [49]; the diffusion coefficient of perchlorate in aqueous solution is around $D = 2 \times 10^{-9}$ m$^2$/s [61]. The diffusion time scale across the polymer layer can be estimated to be $L^2/D = 0.2$ s, which is much shorter than the polarization time even under the fastest scan rate of 200 mV/s (~4 s) used in this study. Thus, we can rule out the possibility of a diffusion limited response embodied in the Randles-Sevcik equation. This study therefore is a strong evidence that for redox-active (Faradaic) materials under high supporting electrolyte concentration and thin film thickness, the charging process is limited by the electron transfer kinetics, but not by the diffusion of either the mobile or electroactive species, although the kinetically accessible surface charge is affected by scan rate due to imperfect electronic conductance of the electrode. The linear dependence of peak current with scan rate was also observed in experimental studies on poly(3,4-ethylenedioxythiophene) (PEDOT)-quinone conducting redox polymers [62], lignin composite films [63] and α-zirconium phosphate nanosheet complexes with polyaniline (PANI) [64]. Here, by incorporating experimental measurements with a physics-based equivalent circuit model, we provide a satisfying explanation for the unique charging and discharging characteristics of redox active electrodes, which clears the confusion between diffusion limited and kinetically limited mechanisms.

The model developed here is easy and generic enough to be used by experimentalists to extract intrinsic parameters of redox active materials from cyclic voltammetry measurements. And the material properties can be further incorporated into a more detailed model to accommodate realistic considerations, including mass transfer limitations [43], variations of ion concentrations in both spatial and time coordinates, which are common in electrosorption applications, including but not limited to deionization, water remediation, energy storage, bio-sensing etc.

## Conclusions

A physics-based equivalent circuit model has been developed that captures the major physical processes of the redox active (Faradaic) materials in electrochemically mediated adsorption applications without loss of simplicity and intuition. A redox polymer, polyvinylferrocene, complexed with carbon nanotubes was studied here as a validation of the model with redox active materials. By incorporating both a voltage dependent capacitance and charge transfer resistance, the simulated results match well with the experimental cyclic voltammetry measurements. The voltage dependent capacitance came from the thermodynamics of the electrosorption process and a Frumkin isotherm (regular solution model) has been shown to describe the process well. The voltage dependent charge transfer resistance can be derived from Butler-Volmer kinetics from non-equilibrium thermodynamics. Cyclic voltammetry experiments are conducted under high concentrations of supporting electrolyte, to allow mass transfer limitations of the mobile ionic species to be negligible. Key parameters of the redox active material can be extracted by fitting the model to the experimental data, such as charge transfer rate, total surface site density, etc. The physics based equivalent circuit model developed here is simple enough to be implemented and extended to the characterization of other redox active systems and provides a quantitatively good understanding of the Faradaically modulated electrosorption materials, and can be a starting point for future integration into more detailed transport models or process level models for process optimization and scale-up. This work should be of interest to researchers in electrochemical water treatment, electrical double layer theory coupled with ion transport and energy storage applications with redox chemistry.

# Figures

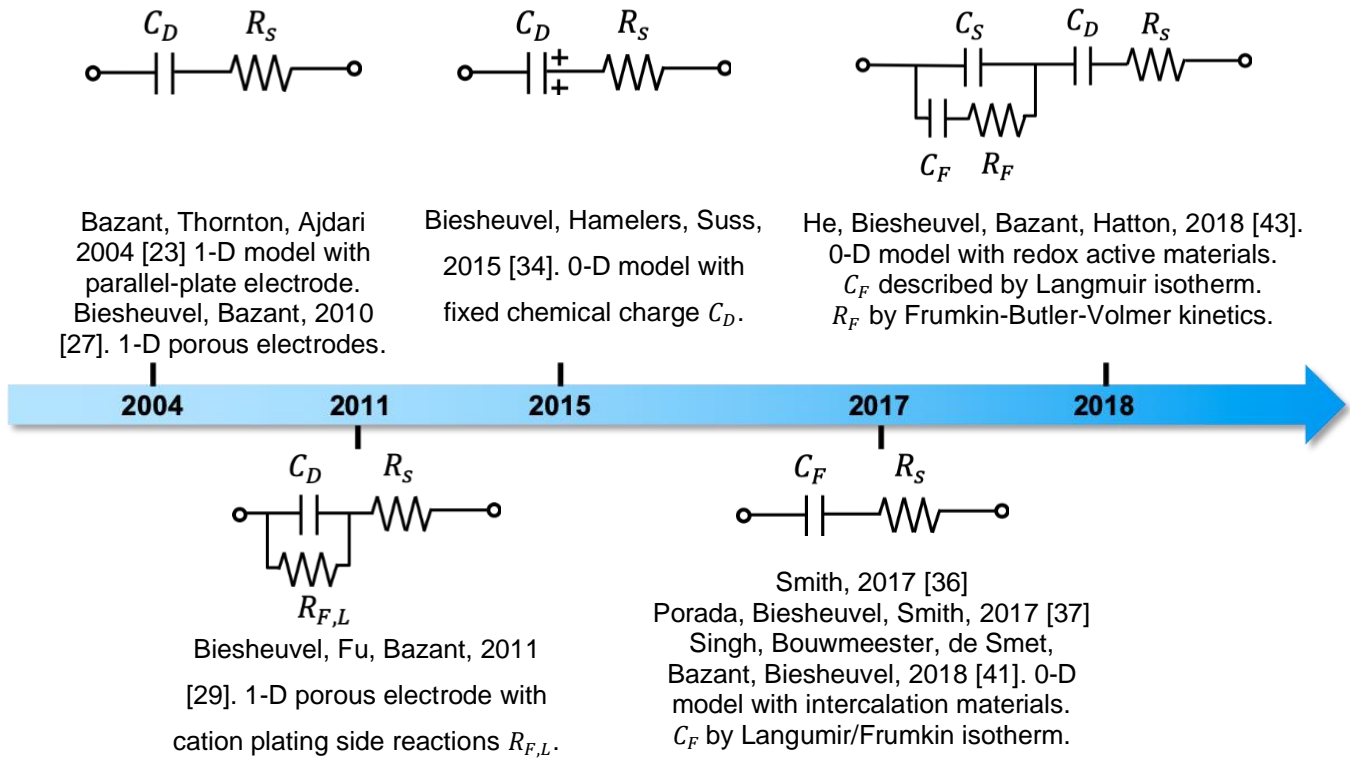

Figure 1. Timeline for mathematical models of (pseudo-)capacitive deionization, represented here as equivalent circuit elements for the interfacial electrochemistry, which may be used in either 0-D models for uniform parallel plate electrodes or 1-D transmission line models for non-uniformly charging porous electrodes. $C_D$ denotes the double layer capacitance (which may contain the diffuse Gouy-Chapman and/or Stern layer), $R_S$ denotes the solution resistance, $C_F$ and $R_F$ denotes the capacitance and charge transfer resistance of the Faradaic reaction.

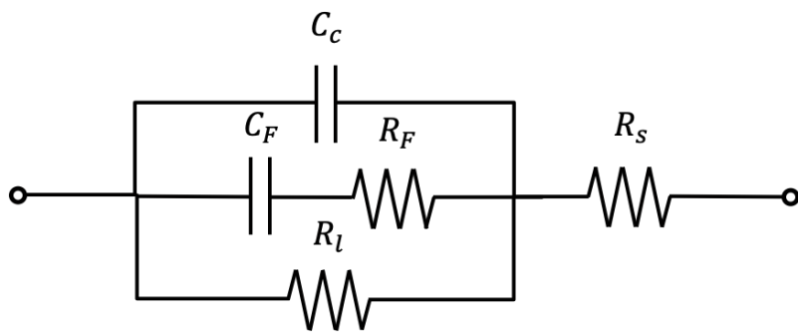

Figure 2. Physics based equivalent circuit model for redox active (Faradaic) electrodes. $C_F$ denotes the Faradaic (pseudo-)capacitance, $R_F$ denotes the Faradaic charge transfer resistance, $R_S$ is the solution resistance, $C_C$ denotes the double layer capacitance, and $R_l$ is the leakage resistance that comes from parasitic reactions.

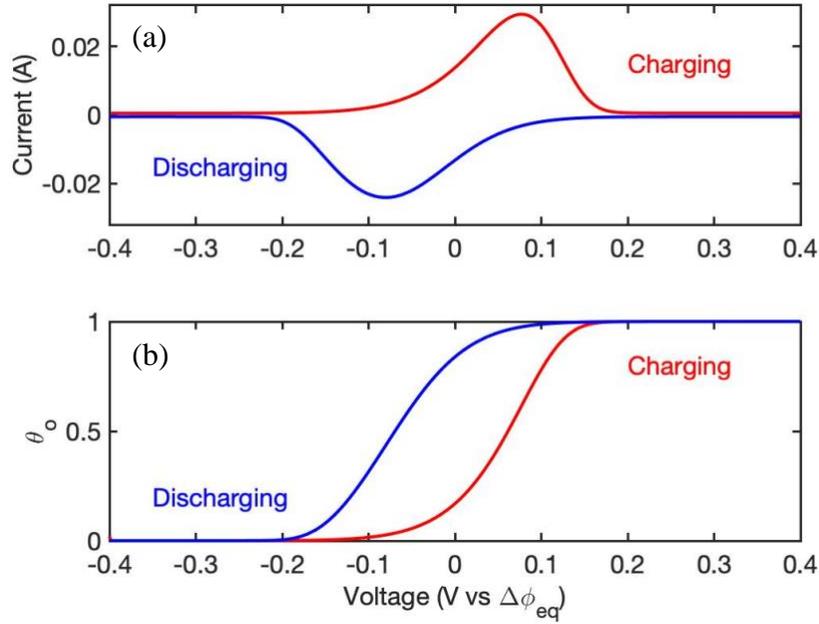

Figure 3. Simulation results of the physics based equivalent circuit model under cyclic voltammetry. Current response (a) and state of charge (b) versus applied voltage. Applied potential has been shifted by the equilibrium potential $\Delta\phi_{eq} = 0.3$ V vs. Ag/AgCl. $k_0 = 0.4$ s$^{-1}$, $\Gamma_s = 4 \times 10^{-3}$ mol/m$^2$, $C_c = 50$ F/m$^2$, $\Omega = -620$ J/mol, $R_s = 0$ Ω, $R_l = 1 \times 10^5$ Ω. Scan rate, $v = 100$ mV/s.

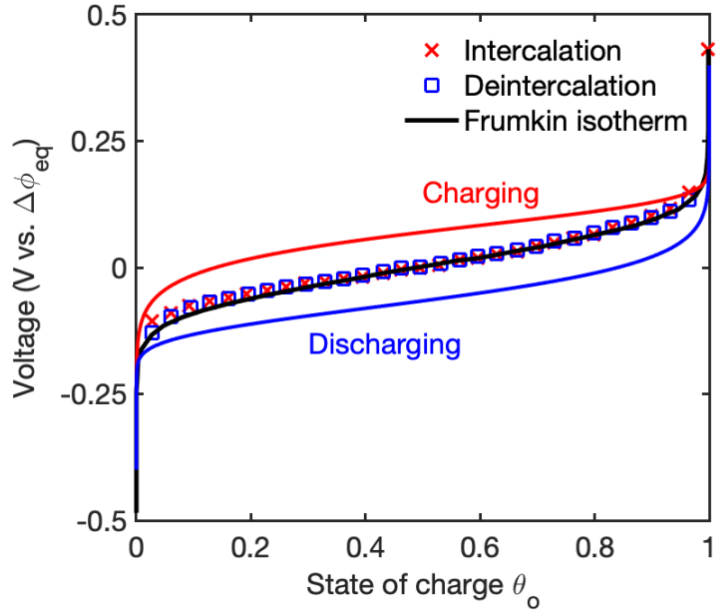

Figure 4. Comparison of the voltage (shifted by the equilibrium voltage $\Delta\phi_{eq}$) versus state of charge (i.e. the fraction of the oxidized state $\theta_O$) at 100 mV/s between the equivalent circuit model and the equilibrium model from the Frumkin isotherm. The red and blue curves are the simulation results of the equivalent circuit model developed in this work. The rests of the figure are redrawn using the extracted data and model from Porada et al [37] with permission. The experimental data (red dots and blue diamonds) were obtained by electrochemical titration of a nickel hexacynoferrate (NiHCF) electrode in 1 M $Na_2SO_4$. The electrode voltage was measured continuously by applying a constant current of 60 mA/g (of NiHCF) for 118 s and then setting the current to zero for 60 s, alternatively. And the charging/discharging curves are constructed using the electrode voltage at the end of the "rest" period, which shows the equilibrium potential vs. electrode charge during the intercalation (red dots) and deintercation (blue diamonds) steps. The black line gives the fitted curve by the Frumkin isotherm, i.e. equation (2). The intercalation degree $\theta$ in reference [37] is equivalent to $1-\theta_O$.

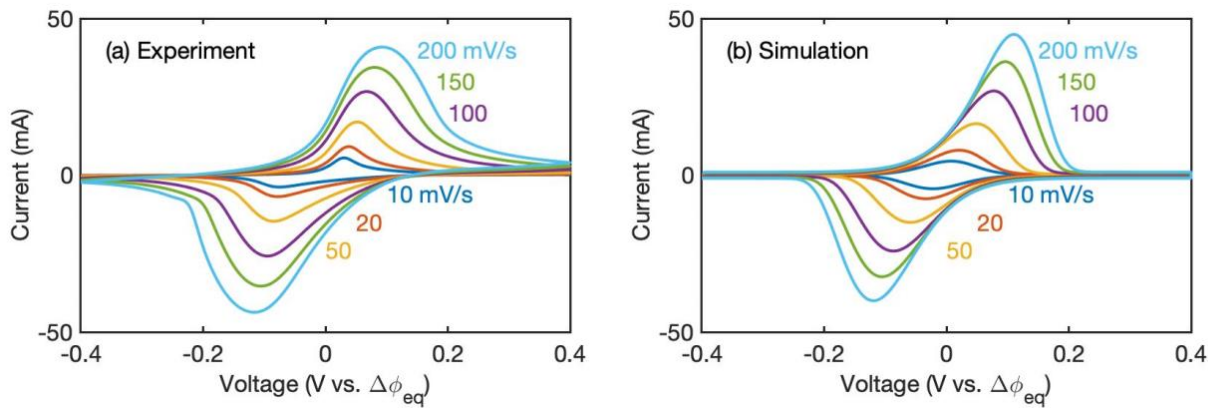

Figure 5. Cyclic voltammetry of (a) experimental (b) simulation results under various scan rates from 10 mV/s to 200 mV/s. Parameter values used for the simulation $k_0 = 0.4 \text{ s}^{-1}$, $C_c = 50 \text{ F/m}^2$, $\Omega = -620 \text{ J/mol}$, $R_s = 0 \text{ }\Omega$, $R_l = 1 \times 10^5 \text{ }\Omega$.

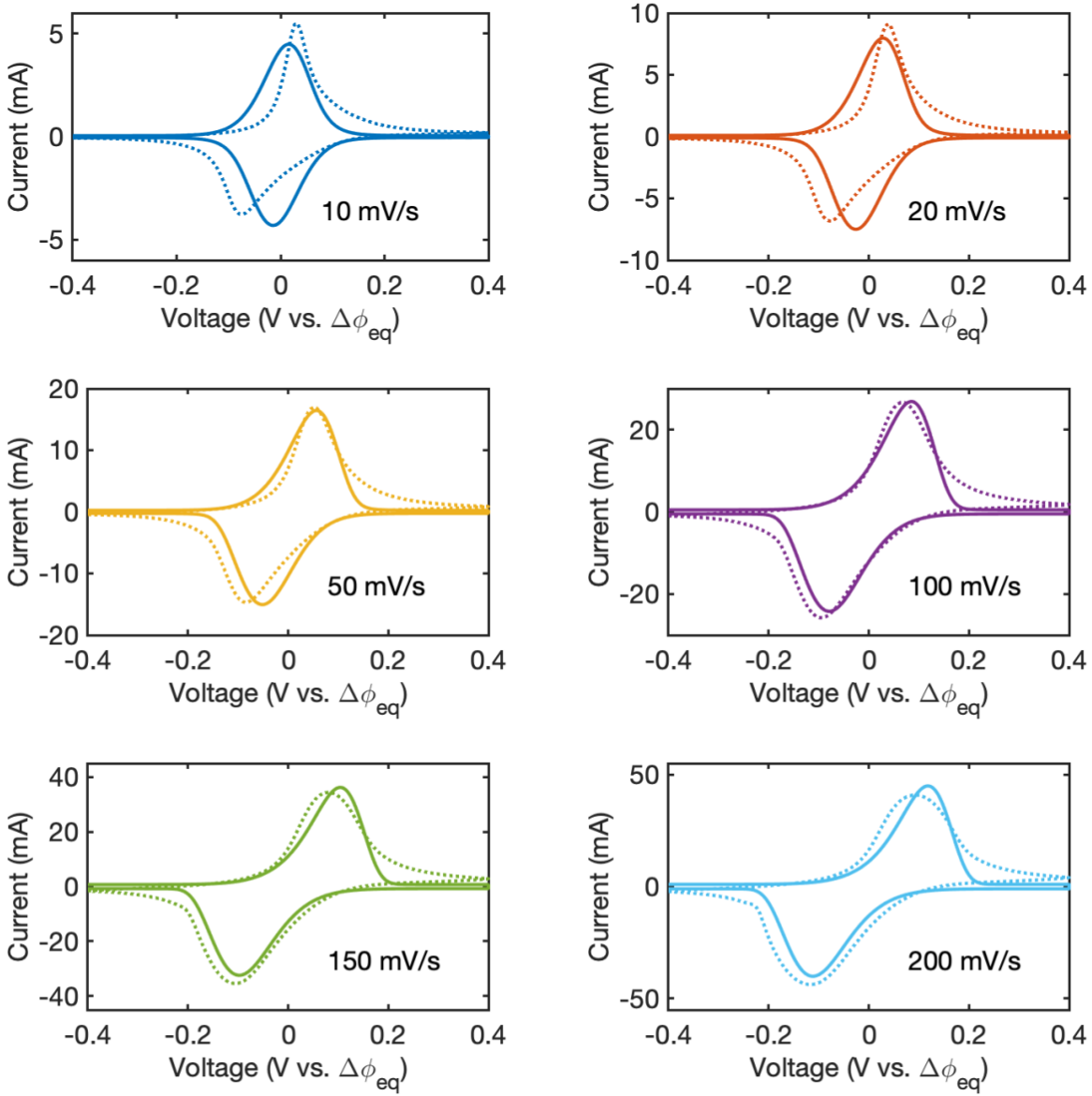

Figure 6. Comparison of experimental data (dotted line) and modeling results (solid line) of cyclic voltammetry under various scan rates from 10 mV/s to 200 mV/s. Parameter values used for the simulation $k_0 = 0.4 \text{ s}^{-1}$, $C_c = 50 \text{ F/m}^2$, $\Omega = -620 \text{ J/mol}$, $R_s = 0 \text{ } \Omega$, $R_l = 1 \times 10^5 \text{ } \Omega$. $\Gamma_s$ varies with scan rate, as shown in Figure 7.

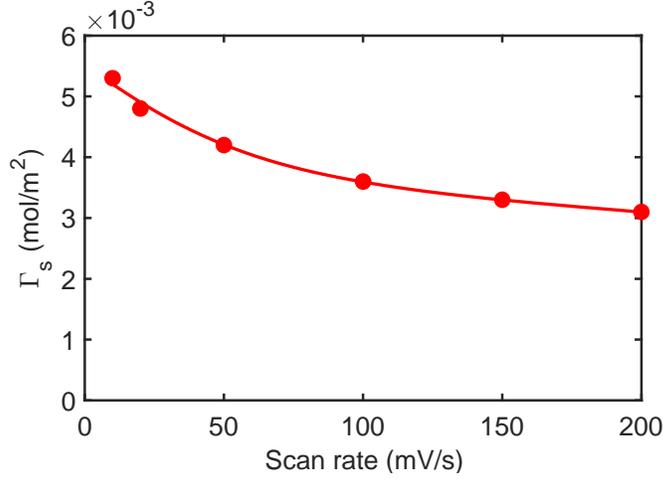

Figure 7. Total accessible surface site density of the redox species $\Gamma_s$ as a function of the scan rate. The dots are the (least-square fitted) surface site density values between the equivalent circuit model and the experimental data and the line is a smooth connection of the dots.

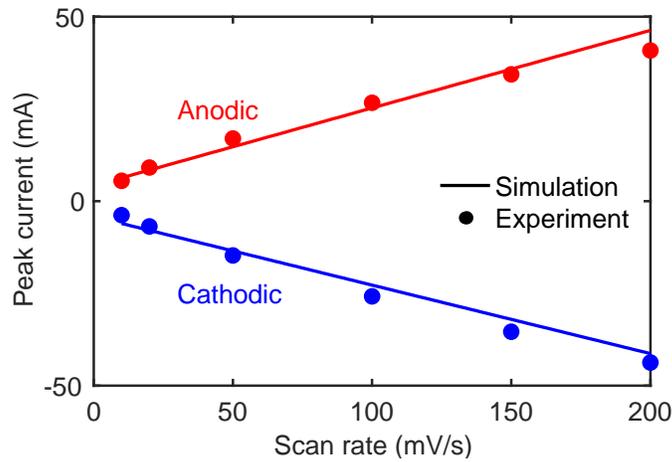

Figure 8. Comparison of experimental and simulation results of the peak currents under various scan rates. The symbols are the experimental data, and the solid line are the peak currents from numerical simulation with the best fitted parameters: $k_0 = 0.4 \text{ s}^{-1}$, $C_c = 50 \text{ F/m}^2$, $\Omega = -620 \text{ J/mol}$, $R_s = 0 \text{ }\Omega$, $R_l = 1 \times 10^5 \text{ }\Omega$. $\Gamma_s$ varies with scan rate, as shown in Figure 7.